\RequirePackage{ifpdf}
\RequirePackage{xspace}
\pdfoutput=1
\ifpdf
  \documentclass[twocolumn,showpacs,aps,prl,amsfonts,amssymb,amsmath,floatfix,superscriptaddress,letterpaper]{revtex4}
  \usepackage{graphicx}
  \usepackage{hyperref}
\else
  \documentclass[twocolumn,showpacs,aps,prl,amsfonts,amssymb,amsmath,floatfix,superscriptaddress,letterpaper]{revtex4}
  \usepackage{graphicx}
  \usepackage[pdfcreator={LaTeX with hyperref package},%
              pdfproducer={dvips + ps2pdf}]{hyperref}
\fi
\usepackage{relsize}

\newcommand{\BABARPubYear}   {08}
\newcommand{\BABARPubNumber} {002}
\newcommand{\SLACPubNumber}  {13217}
\newcommand{\LANLNumber}     {0805.0497}

\graphicspath{{./}}
\hypersetup{%
  pdftitle={Measurement of the Mass Difference m(B0)-m(B+)},%
  pdfauthor={B. Aubert et al. (BABAR Collaboration)},%
  pdfsubject={BABAR-PUB-\BABARPubYear/\BABARPubNumber,
 SLAC-PUB-\SLACPubNumber, arXiv:\LANLNumber\ [hep-ex]},%
  pdfkeywords={BABAR, BABAR-PUB-\BABARPubYear/\BABARPubNumber,
 SLAC-PUB-\SLACPubNumber, arXiv:\LANLNumber\ [hep-ex]},%
  bookmarks=true,%
  bookmarksnumbered=true,%
  bookmarksopen=false,%
  pdfpagemode=UseOutlines,%
  pdfstartview=Fit,%
  pdfpagelayout=SinglePage,%
  hypertexnames=false,%
  breaklinks=true,%
  colorlinks=true,%
  linkcolor={black},%
  citecolor={black},%
  filecolor={black},%
  urlcolor={black}%
}

\newcommand{\reffig}   [1] {Figure~\ref{fig:#1}}
\newcommand{\reftab}   [1] {Table~\ref{tab:#1}}
\newcommand{\refeq}    [1] {Eq.~\ref{eq:#1}}
\newcommand{\qrkbar}   [1] {\kern 0.05em\overline{\kern -0.05em #1}{}\xspace}
\newcommand{\qpair}    [1] {\ensuremath{#1\qrkbar{#1}}\xspace}
\def\qqq                   {\qpair{q}}

\def\pmtab                 {\ensuremath{\,\pm\,}\xspace}
\def\tabT                  {\rule{0pt}{2.6ex}}
\def\tabB                  {\rule[-1.2ex]{0pt}{0pt}}
\def\rms                   {{rms}{}\xspace}
\def\babar                 {\mbox{\slshape B\kern-0.1em{\smaller A}\kern-0.1em
   B\kern-0.1em{\smaller A\kern-0.2em R}}}
\def\epem                  {\ensuremath{e^+e^-}\xspace}
\def\mumu                  {\ensuremath{\mu^+\mu^-}\xspace}
\def\pip                   {\ensuremath{\pi^+}\xspace}
\def\pim                   {\ensuremath{\pi^-}\xspace}
\def\Kp                    {\ensuremath{K^+}\xspace}
\def\Kz                    {\ensuremath{K^0}\xspace}
\def\Kstarz                {\ensuremath{K^{*0}}\xspace}
\def\jpsi                  {\ensuremath{{J\mskip -3mu/\mskip -2mu\psi\mskip
   2mu}}\xspace}
\def\B                     {\ensuremath{B}\xspace}
\def\Bbar                  {\kern 0.18em\overline{\kern -0.18em B}{}\xspace}

\def\BB                    {\ensuremath{B\Bbar}\xspace} 
\def\Bz                    {\ensuremath{B^0}\xspace}
\def\Bzb                   {\ensuremath{\Bbar^0}\xspace}
\def\BzBzb                 {\ensuremath{\Bz {\kern -0.16em \Bzb}}\xspace}
\def\Bp                    {\ensuremath{B^+}\xspace}
\def\Bm                    {\ensuremath{B^-}\xspace}
\def\BpBm                  {\ensuremath{\Bp {\kern -0.16em \Bm}}\xspace}
\def\to                    {\ensuremath{\rightarrow}\xspace}
\newcommand{\mev}          {\ensuremath{\mathrm{\,Me\kern -0.1em V}}\xspace}
\newcommand{\gevc}         {\ensuremath{{\mathrm{\,Ge\kern -0.1em V\!/}c}}\xspace}
\newcommand{\mevc}         {\ensuremath{{\mathrm{\,Me\kern -0.1em V\!/}c}}\xspace}
\newcommand{\gevcc}        {\ensuremath{{\mathrm{\,Ge\kern -0.1em V\!/}c^2}}\xspace}
\newcommand{\mevcc}        {\ensuremath{{\mathrm{\,Me\kern -0.1em V\!/}c^2}}\xspace}
\def\invfb                 {\ensuremath{\mbox{\,fb}^{-1}}\xspace}
\def\mrad                  {\ensuremath{\rm \,mrad}\xspace}
\def\gsim                  {{~\raise.15em\hbox{$>$}\kern-.85em
   \lower.35em\hbox{$\sim$}~}\xspace}
\def\lsim                  {{~\raise.15em\hbox{$<$}\kern-.85em
   \lower.35em\hbox{$\sim$}~}\xspace}
\mathchardef\Upsilon="7107
\def\Y#1S{\ensuremath{\Upsilon{(#1S)}}\xspace}%
\def\pep2{PEP-II}%

\begin{document}

\preprint{\babar-PUB-\BABARPubYear/\BABARPubNumber} 
\preprint{SLAC-PUB-\SLACPubNumber} 
\begin{flushleft}
\babar-PUB-\BABARPubYear/\BABARPubNumber\\
SLAC-PUB-\SLACPubNumber\\[8mm]%[2mm]%
%arXiv:\LANLNumber\ [hep-ex]\\[8mm]
\end{flushleft}

\title{{\large \bf\boldmath%
Measurement of the Mass Difference $m(\Bz)-m(\Bp)$}}

\author{B.~Aubert}
\author{M.~Bona}
\author{Y.~Karyotakis}
\author{J.~P.~Lees}
\author{V.~Poireau}
\author{E.~Prencipe}
\author{X.~Prudent}
\author{V.~Tisserand}
\affiliation{Laboratoire de Physique des Particules, IN2P3/CNRS et Universit\'e de Savoie, F-74941 Annecy-Le-Vieux, France }
\author{J.~Garra~Tico}
\author{E.~Grauges}
\affiliation{Universitat de Barcelona, Facultat de Fisica, Departament ECM, E-08028 Barcelona, Spain }
% ...
\collaboration{The \babar\ Collaboration}
\noaffiliation
\date{\today}

\begin{abstract}
Using $230\times 10^6$ \BB\ events recorded with the \babar\ detector at the
\epem\ storage rings \pep2, we reconstruct approximately 4100
$\Bz\to\jpsi\Kp\pim$ and 9930 $\Bp\to\jpsi\Kp$ decays with $\jpsi\to\mumu$ and
\epem. From the measured \B-momentum distributions in the \epem\ rest frame, we
determine the mass difference $m(\Bz)-m(\Bp)=(+0.33\pm 0.05\pm 0.03)\mevcc$.
\end{abstract}


\maketitle

Mass differences $\Delta m_M=m(M^0)-m(M^+)$ probe the size of Coulomb
contributions to the quark structure of pseudoscalar mesons $M$. The values of
$\Delta m_M$ for $\pi$, $K$, and $D$ mesons are experimentally well known; in
units of $\mevcc$ they are $\Delta m_\pi=-4.5936\pm 0.0005$,
$\Delta m_K=+3.97\pm 0.03$, and $\Delta m_D=-4.78\pm 0.10$~\cite{Yao:2006px}.
For \B mesons, $\Delta m_\B=(+0.37\pm 0.24)\mevcc$~\cite{PDGweb} is less precise
and compatible with zero. Quark-model calculations~\cite{Goity:2007fu} give
$\Delta m_\B$ near $+0.3\mevcc$ but are quite uncertain since the contributions
from the quark-mass difference $m(d)-m(u)$ and from the Coulomb effects have
similar magnitudes and opposite signs. In the case of $\Delta m_D$, the two
contributions enter with the same signs.

The value of $\Delta m_\B$ is an important input for estimating the decay ratio
$R=\Gamma[\Y4S\to\BpBm]/\Gamma[\Y4S\to\BzBzb]$ which in turn is essential for
determining \Bp and \Bz decay fractions at \epem\ colliders where \B mesons are
produced in decays of the \Y4S. The leading contribution to $R$ is given by the
vector nature of the matrix element and by kinematics; at fixed energy it is
\begin{equation}
R_0 = \left[p^*(\Bp)/p^*(\Bz)\right]^3\approx 1+3 m_\B \Delta m_\B/p^{*2}_\B\,,
\label{eq:eq1}
\end{equation}
where $p^*(\Bp)$ and $p^*(\Bz)$ are the \Bp and \Bz momenta in the
center-of-mass system (cms) at this energy, and $p^*_\B$ and $m_\B$ are the mean
values of the two momenta and masses, respectively. For $|\Delta m_\B|$ below
0.5\mevcc, the quark structures of \Y4S and \B mesons and the Coulomb
interaction~\cite{Dubynskiy:2007xw} may lead to $|R-R_0|>R_0-1$.

For measuring $\Delta m_\B$, we use 210\invfb of \epem\ annihilation data
recorded on the \Y4S resonance with the \babar\ detector~\cite{Aubert:2001tu} at
the SLAC \epem\ storage rings \pep2~\cite{pep:1993mp}. Charged-particle momenta
are measured by the tracking system consisting of a five-layer double-sided
silicon vertex tracker and a 40-layer drift chamber, both located in a
$1.5\,{\rm T}$ magnetic field of a superconducting solenoid. Transverse momenta
$p_T$ are determined with a resolution of about
$\sigma(p_T)/p_T=0.0013\times p_T c/{\rm Ge\kern -0.1em V}+0.0045$ and track
angles with resolutions around 0.4\mrad.

The \B mesons are reconstructed in two decay modes with low background level:
$\Bz\to\jpsi\Kp\pim$ and $\Bp\to\jpsi\Kp$~\cite{useCC}, where $\jpsi\to\mumu$ or
\epem in both modes. Measurements of \Kz and \jpsi invariant masses show that
relative momentum uncertainties $\delta p/p$, originating from the limited
knowledge of the magnetic field and the charged-particle energy losses, are
below $4\times 10^{-4}$. A momentum uncertainty of this size leads to \B-meson
mass uncertainties of the order of 1\mevcc. The mass difference $\Delta m_\B$
can be determined with much higher precision using \B-meson momenta because the
decay $\Y4S\to\BB$ produces \B mesons with low momenta,
$p^*(\B)\approx 320\mevc$. At fixed cms energy $\sqrt{s}$ we have
\begin{equation}
m^2(\Bp)c^2 + p^{*2}(\Bp) = m^2(\Bz)c^2 + p^{*2}(\Bz)\,,
\label{eq:eq2}
\end{equation}
\begin{equation}
\Delta m_\B = -\Delta p^*\times
              \frac{p^*(\Bz) + p^*(\Bp)}{\left[m(\Bz) + m(\Bp)\right]c^2}
\label{eq:eq3}
\end{equation}
where $\Delta p^*=p^*(\Bz)-p^*(\Bp)$. The track-momentum uncertainties lead to
$\delta p^*<4\times 10^{-4}\times 320\mevc$,
$\delta(\Delta p^*)\lsim\sqrt{2}\times\delta p^*$, and, using \refeq{eq3},
$\delta(\Delta m_\B)<0.01\mevcc$ which is two orders of magnitude smaller than
the 1\mevcc estimate using invariant masses.

The energy spread of the \pep2\ beams gives a $\sqrt{s}$ distribution with a
\rms width of about 5\mev, resulting in broad distributions of the true momenta
$p_{\rm true}^*(\B)$ with \rms widths of about 40\mevc. The reconstructed $p^*$
spectra are only slightly wider since the detector resolution is only
$\sigma(p^*-p_{\rm true}^*)\approx 15\mevc$ in the selected \B-decay modes,
where $\sigma$ is the \rms width. As input for \refeq{eq3}, we use the mean
values ${\hat p}^*(\Bz)$ and ${\hat p}^*(\Bp)$ of the reconstructed $p^*$
spectra. The presence of background prevents the two ${\hat p}^*$ values from
being obtained as algebraic means of all measured $p^*$ values. Instead, they
are determined from fits with analytic functions for the signal and the
background shapes.

The size of a possible bias from the mean-$p^*$ method is estimated by Monte
Carlo (MC) simulations in two steps. The influence of the beam smearing and the
\Y4S line shape is studied by determining the means ${\hat p}_{\rm true}^*(\B)$
using a MC simulation in the cms without detector. We use a Gaussian cms-energy
distribution with two parameters: $\sqrt{s}_{\rm mean}$ and $\sigma_{\sqrt{s}}$,
and $\Y4S\to\BB$ line shapes with four parameters: $m(\Y4S)$, $m(\Bz)$,
$m(\Bp)$, and $\Gamma_0$, where the latter is the total width at $s=m^2(\Y4S)$.
The line shape is parametrized following ref.~\cite{Aubert:2004pwa}; it includes
initial state radiation, a relativistic Breit-Wigner function with
energy-dependent width, $m(\B)$- and $s$-dependent phase space factors, and
meson-structure effects. Because of the $m(\B)$ dependence of the phase space
factor, the line shapes differ for \Bz and \Bp. We fix $\Delta m_\B$ to either
$+0.3$ or $+0.4\mevcc$ and vary the other parameters in the range of the results
of ref.~\cite{Aubert:2004pwa}. We determine ${\hat p}_{\rm true}^*(\Bz)$ and
${\hat p}_{\rm true}^*(\Bp)$ for each set of parameters and find that the
derived $\Delta m_\B$ results from \refeq{eq3} are equal to the MC input within
$\pm 2\%$. The \rms widths $\sigma_{p^*}$ of the two $p_{\rm true}^*$
distributions are found to be different in agreement with
\begin{equation}
\sigma_{p^*}(\Bp) /\sigma_{p^*}(\Bz) = {\hat p}^*(\Bz)/{\hat p}^*(\Bp)\,,
\label{eq:eq4}
\end{equation}
as simple consequence of \refeq{eq2}.

The detector influence on the $\Delta m_\B$ bias is studied by a full MC
simulation of generic \BB decays with GEANT4~\cite{Agostinelli:2002hh}. The
simulation includes all detector and reconstruction effects and the same
Gaussian cms-energy distribution as above, but uses a simpler $\Y4S\to\BB$ line
shape with fixed $\Gamma(s)$, without initial state radiation, and without
meson-structure effects. The results on the means of $p^*-p_{\rm true}^*$ are
given in the discussion of the systematic uncertainties.

\begin{figure*}
\begin{center}
\includegraphics[width=\textwidth]{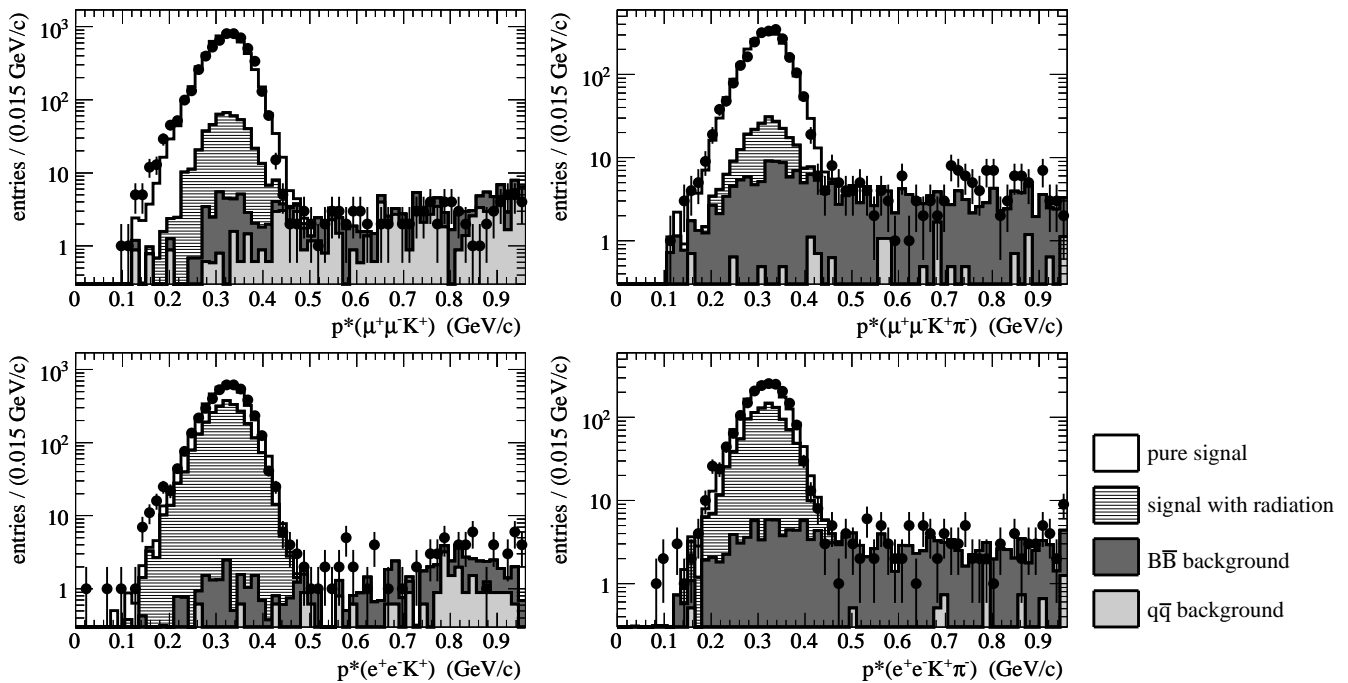}
\end{center}
\caption{\label{fig:bMomCM}%
The $p^*$ distributions of the selected data (dots with error bars) and of the
selected MC events (histograms stacked on top of each other). The left side is
for \Bp, the right side for \Bz, the top for $\jpsi\to\mu\mu$, the bottom for
$\jpsi\to ee$.}
\end{figure*}

The same GEANT4-based MC simulation is used to determine the selection criteria
for \B reconstruction and to find the fit-function types for signal and
background in the reconstructed $p^*$ spectra. The \jpsi decays into $\mu\mu$
and $ee$ are studied separately in order to control the influences of
bremsstrahlung in the $ee$ channel, simulated by PHOTOS~\cite{Barberio:1993qi}
and GEANT4. Muons are identified using a neural network with a high efficiency
of 0.90 accepting a rather high probability for pion misidentification (misid)
of 0.08, while electrons are identified using a likelihood selector with an
efficiency of 0.95 and a pion-misid probability of $10^{-3}$. Electron tracks
are combined with up to three nearby photons into electron candidates using a
bremsstrahlung-recovery algorithm. Pairs of electrons or muons with opposite
charge are fitted to a common vertex. All pairs with a vertex fit probability
$P>10^{-4}$ and an invariant mass between 3.057 and 3.137\gevcc are selected as
\jpsi candidates. Because of background from two pions in jet-like
$\epem\to\qqq$ events, we also require $|\cos(\theta_H)|<0.9$ for
$\jpsi\to\mu\mu$ candidates, where $\theta_H$ is the angle between one muon and
the \B candidate in the \jpsi rest frame. Since the pion misid is much lower for
electrons, this cut is not applied in the $\jpsi\to ee$ mode. In the \Bz mode
with $\jpsi\to\mu\mu$ we require in addition that the normalized second
Fox-Wolfram moment $R_2$~\cite{Fox:1978vw} of the event is less than 0.4. 

Charged kaons are identified using a likelihood selector, based on the DIRC
system~\cite{Schwiening:2005bv} of \babar, with an efficiency of 0.95 and a
pion-misid probability of 0.05. The $K\pi$ pairs are formed from two oppositely
charged tracks, one identified as a kaon and the other as a pion; the fit to a
common vertex must give a fit probability $P>10^{-4}$. For suppressing
background, we require an invariant mass $m(K\pi)=m(\Kstarz)\pm 75\mevcc$. The
\Bz and \Bp candidates are formed by combining the \jpsi with the $K\pi$-pair
candidates and with charged tracks identified as kaons, respectively. We also
require a fit probability $P>10^{-4}$ for the common vertex. The \B candidates
are further selected by their value of $\Delta E^*=E_\B^*-\sqrt{s}/2$, where
$E_\B^*$ is the energy of the \B candidate in the cms.

To optimize signal versus background in the $p^*$ distributions and to account
for bremsstrahlung, we have chosen four different $\Delta E^*$ selection
criteria. For $\jpsi\to\mu\mu$ we choose $|\Delta E^*|<55\mev$ for the \Bp and
$|\Delta E^*|<25\mev$ for the \Bz. For $\jpsi\to ee$ we take
$-60<\Delta E^*<50\mev$ and $-30<\Delta E^*<20\mev$ in \Bp and \Bz decays
respectively. For the \Bp this corresponds to $\pm 3$ \rms widths of the signal,
for the \Bz to $\pm 1.5$ \rms. The tighter criteria in \Bz decays, where the
background is an important contribution to the final systematic uncertainty on
$\Delta m_\B$, are justified by the negligible correlations between $\Delta E^*$
and $p^*$ and by the MC validation as described below. After applying the
$\Delta E^*$ criteria to the \B candidates, there remain events with more than
one candidate. The fraction is negligible for \Bp (0.10\% of all events) but is
1.5\% for \Bz. If there are multiple \B candidates in the event, we choose the
one with the best \B-vertex fit. The selection criteria for data and MC events
are identical with one exception: In the data, because of a bias in the \jpsi
mass reconstruction owing to track-momentum uncertainties, the lower and upper
limits for $m(\mu\mu)$ and $m(ee)$ are shifted by $-2\mevcc$.

\reffig{bMomCM} shows the $p^*$ distributions of the selected data and those
from the MC simulation. The MC distributions are normalized to the data between
0.12 and 0.45\gevc. They contain contributions from four classes,
\newcounter{cntmc}
\begin{list}{Class \arabic{cntmc},}{%
  \usecounter{cntmc}%
  \setlength{\topsep}{0pt}%
  \setlength{\parsep}{0pt}%
  \setlength{\itemsep}{0pt}%
  \setlength{\leftmargin}{1.5em}%
  \setlength{\rightmargin}{0pt}%
  \setlength{\itemindent}{0pt}%
  \setlength{\labelwidth}{1.1em}%
  \setlength{\labelsep}{0.4em}%
}
\item\label{it:mc1} ``pure signal'', candidates where all tracks originate from
true \B-decay particles into the given mode and where the decays contain no
photons including those combined into electron candidates with the
bremsstrahlung-recovery algorithm,
\item\label{it:mc2} ``signal with radiation'', like pure signal, but with at
least one photon from bremsstrahlung generated by PHOTOS or GEANT4,
\item\label{it:mc3} ``\BB background'', candidates from \B decays other than
from classes \ref{it:mc1} or \ref{it:mc2}, and
\item\label{it:mc4} ``\qqq\ background'', candidates from non-\BB events.
\end{list}
The third class also contains some signal events with wrong matching of
reconstructed and generated tracks. As can clearly be seen, the \BB background
in \Bz decays is larger than in \Bp decays and the fraction of candidates with
bremsstrahlung is larger in the $\jpsi\to ee$ than in the $\jpsi\to\mu\mu$ mode.
Note that, in spite of observed differences in the invariant lepton-pair mass
and in the $\Delta E^*$ distributions for $ee$ and $\mu\mu$, there is almost no
difference in the shape of the $p^*$ distributions. In \reffig{bMomCM},
differences between data and simulation are seen on both edges of the signal
peaks. They may arise from imperfections in describing the beam energy spread
and the \Y4S line shape which influence \Bz and \Bp decays equally; the
following data analysis has to account for the imperfections.

The mean values of the four $p^*(\B)$ spectra are obtained from fits. The form
of the fit functions is obtained from the MC spectra for ``pure signal'' and the
sum of \BB\ and \qqq\ backgrounds separately. For the signal, we find that a
double-Gaussian function $S(p^*)$ with six parameters is adequate. Its
parameters are: the number $N$ of signal events (sum of classes \ref{it:mc1} and
\ref{it:mc2}), the mean ${\hat p}^*$ and the \rms width $\sigma_{p^*}$ of
$S(p^*)$, the fraction $f$ of the subdominant Gaussian function, the
peak-position difference $\Delta$ and the width ratio $r_\sigma$ of the two
Gaussian functions. The $\chi^2$ fits of $S(p^*)$ to the ``pure signal''
contributions in the four spectra of \reffig{bMomCM} are of good quality. The
fit-parameter values are similar in all four spectra; only the $\sigma_{p^*}$
values are slightly larger in $ee$ than in $\mu\mu$ decays. It has been checked
that $S(p^*)$ with the same parameters as for ``pure signal'' also describes the
$p^*$ distributions of ``signal with radiation'' for both \Bz and \Bp decays.

The backgrounds for \Bz and \Bp are very different, requiring two different
function types $U_0(p^*)$ and $U_+(p^*)$. We find that polynomials are adequate
in both cases, linear for \Bp and of fifth degree for \Bz. The polynomials are
determined by fits to the sum of the MC background histograms. Because of the
complication with the mismatched signal MC events, we have to use fit functions
$U_{0,+}(p^*)+S(p^*)$ for determining the background polynomials, where $S(p^*)$
is the best-fit signal function with free normalization.

In the fits of $S(p^*)+U_{0,+}(p^*)$ to the $p^*$ distributions of real data, we
choose binned maximum-likelihood fits between 0.12 and 0.95\gevc with bin widths
of 0.015\gevc. The background polynomials are used with free normalizations
$r_{\rm bg}$ but with shape parameters as given by the MC fits. In the signal
function, all six parameters are left free since the signal shapes differ in
data and MC because of the imperfect MC simulation. Since the $\sqrt{s}$
spectrum dominates the shapes of the two $p^*$ spectra, the parameters $f$,
$\Delta$, and $r_\sigma$ are constrained to be equal for \Bz and \Bp.

Before fitting the real data, we apply the fit to $p^*$ distributions of the MC
simulation. We divide the sample of reconstructed MC events in five parts of
equal size, each with the same integrated luminosity as the data. The 10 fit
results, combining $\jpsi\to\mu\mu$ and $ee$, have a mean of
$\Delta p^*={\hat p}^*(\Bz)-{\hat p}^*(\Bp)=-4.7\mevc$ with a \rms of 0.4\mevc
in good agreement with the MC input of $-5.1\mevc$.

\begin{figure*}
\begin{center}
\includegraphics[width=\textwidth]{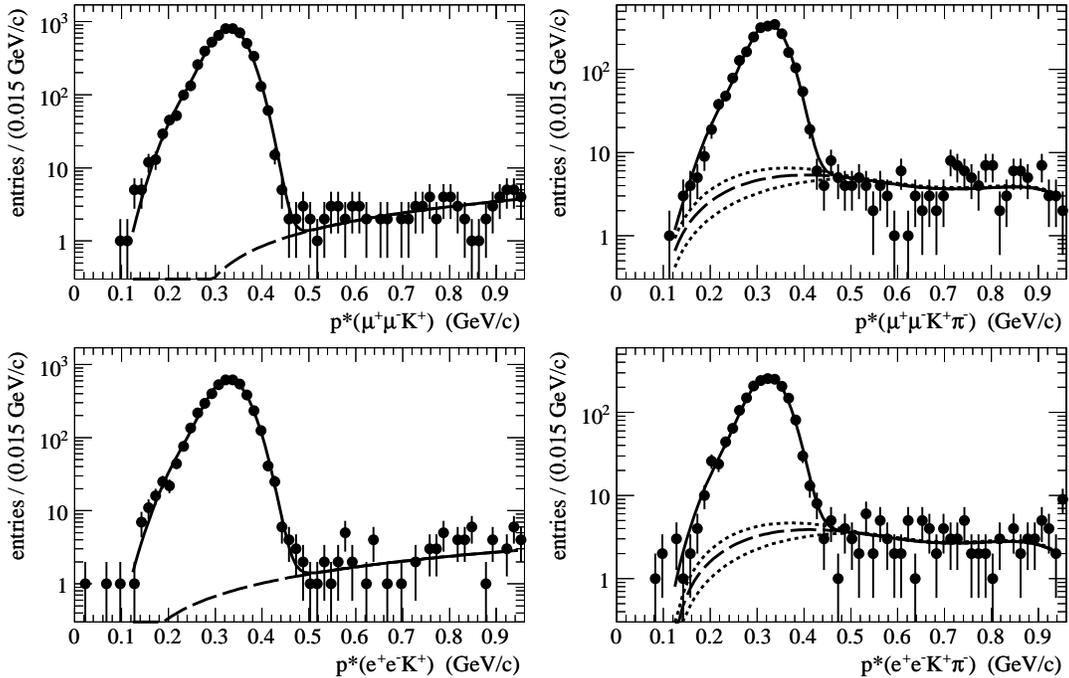}
\end{center}
\caption{\label{fig:bMomCMData}%
The fitted $p^*$ distributions in data, left side for \Bp, right side for \Bz,
top for $\jpsi\to\mu\mu$, bottom for $\jpsi\to ee$. The dashed lines show the
background polynomials; the dotted lines for \Bz show the changed backgrounds
for the systematic-uncertainty estimate.}
\end{figure*}

\reffig{bMomCMData} shows the $p^*$ distributions of the selected data events
together with the best-fit functions. The fit results are given in
\reftab{SigFitData}, and the derived $\Delta p^*$ values are
$(-4.8\pm 1.1)\mevc$ for the $\jpsi\to\mu\mu$ and $(-6.4\pm 1.3)\mevc$ for the
$\jpsi\to ee$ mode. The two values are consistent, we therefore use the weighted
mean
\begin{equation*}
\Delta p^* = (-5.5\pm 0.8 )\mevc
\end{equation*}
for the final result. Before converting $\Delta p^*$ into $\Delta m_\B$, we
present a number of cross-checks and the estimates of all contributions to the
systematic uncertainty.

\begin{table}
\caption{\label{tab:SigFitData}%
Results for fitting the sum of signal and background functions to the four data
$p^*$ spectra in \reffig{bMomCMData}. The parameter $N$ is the number of signal
events and $r_{\rm bg}$ is the ratio of the observed to the simulated background
level. The values for ${\hat p}^*$, $\sigma_{p^*}$, and $\Delta$ are in \mevc.}
\begin{center}
\begin{tabular}{lr@{\pmtab\/}lr@{\pmtab\/}lr@{\pmtab\/}lr@{\pmtab\/}l}
\toprule
& \multicolumn{2}{c}{\tabT\tabB $\Bz,\,\mu\mu$} & \multicolumn{2}{c}{$\Bp,\,\mu\mu$} &
\multicolumn{2}{c}{$\Bz,\,ee$} & \multicolumn{2}{c}{$\Bp,\,ee$} \\
\hline
\tabT $N$          &  $2280$ &   $50$ &  $5580$ &   $70$ &  $1820$ &   $40$ &  $4350$ &   $70$ \\
${\hat p}^*$       & $316.8$ &  $0.9$ & $321.6$ &  $0.6$ & $314.7$ &  $1.1$ & $321.1$ &  $0.7$ \\
$\sigma_{p^*}$     &  $43.0$ &  $0.8$ &  $44.4$ &  $0.5$ &  $44.3$ &  $0.9$ &  $45.4$ &  $0.6$ \\
$f$                & \multicolumn{2}{r@{}}{$0.79$} & \multicolumn{2}{@{}l}{$\pmtab 0.04$} &
\multicolumn{2}{r@{}}{$0.78$} & \multicolumn{2}{@{}l}{$\pmtab 0.06$} \\
$\Delta$           & \multicolumn{2}{r@{}}{$-51$}  & \multicolumn{2}{@{}l}{$\pmtab 7$}    &
\multicolumn{2}{r@{}}{$-48 $} & \multicolumn{2}{@{}l}{$\pmtab 10$}   \\
$r_\sigma$         & \multicolumn{2}{r@{}}{$1.46$} & \multicolumn{2}{@{}l}{$\pmtab 0.08$} &
\multicolumn{2}{r@{}}{$1.48$} & \multicolumn{2}{@{}l}{$\pmtab 0.08$} \\
\tabB $r_{\rm bg}$ &  $1.16$ & $0.10$ &  $1.01$ & $0.11$ &  $1.08$ & $0.11$ &  $1.93$ & $0.24$ \\
\hline\hline
\end{tabular}
\end{center}
\end{table}

The $\Delta p^*$ results from different run periods of the experiment are in
agreement with each other and no charge dependence is observed. We find
${\hat p}^*(\Bp)-{\hat p}^*(\Bm)=-0.3\pm 1.3$ $(-1.0\pm 1.4)\mevc$ for the
$\mu\mu$ ($ee$) mode and ${\hat p}^*(\Bz)-{\hat p}^*(\Bzb)=0.4\pm 1.8$
$(-1.2\pm 2.1)\mevc$ for $\mu\mu$ ($ee$). Varying the $\Delta E^*$ requirements
for the \B candidates by factors of 1.4 up or down changes the central value of
the $\Delta p^*$ result by less than half a standard deviation. No sizable
effect on the central value is seen when removing the requirement on the muon
angle $\theta_H$ for the \jpsi candidates, on $m(K\pi)$, or on the event-shape
parameter $R_2$.

The contributions to the systematic uncertainty of $\Delta p^*$ are summarized
in \reftab{syst}. The influence of the chosen parametrization for the signal
fit-function is estimated by using modified parametrizations. First, we allow
$f$, $\Delta$, and $r_\sigma$ to be different for \Bz and \Bp which results in
$\Delta p^*=(-5.4\pm 0.8)\mevc$. Second, we use one parameter less than in the
nominal fit requiring
$\sigma_{p^*}(\Bz)=\sigma_{p^*}(\Bp)\times{\hat p}^*(\Bp)/{\hat p}^*(\Bz)$ from
\refeq{eq4} resulting in $(-5.7\pm 0.8)\mevc$. We use the observed average
variation of the three fit-method results in data and in the five MC validation
subsamples as an estimate of the systematic uncertainty for the signal
fit-function. Since the backgrounds are small, we also determine ${\hat p}^*$ as
algebraic means of the four $p^*$ spectra between 0.12 and 0.45\gevc after
subtracting the best-fit background functions. The results agree with those in
\reftab{SigFitData} within $\pm 0.1\mevc$ except for ${\hat p}^*(\Bz,\,ee)$
where it is 0.3\mevc lower.

The influence of the background in the $p^*(\Bz)$ spectrum requires special care
and was investigated by three methods. First, we compare the fit results for
various $\Delta E^*$ cuts with $r_{\rm bg}$ free and $r_{\rm bg}=1$. Second, in
order to control the influence of a slightly different background shape in the
signal region, we fit the \Bz data using modified functions ${\tilde U}_0(p^*)$
with the arbitrary shapes of the two dotted lines in \reffig{bMomCMData}. Third,
we select wrong-sign candidates in the channel $\jpsi\Kp\pip$ with all selection
criteria as for the nominal \Bz candidates including those for $m(K\pi)$. The
ratio $Q$ of selected data and MC events is well approximated by the linear
function $Q=0.30+0.78\times p^* c/{\rm Ge\kern -0.1em V}$. The function
$Q\times U_0(p^*)$ is then fitted to the selected \Bz data with $r_{\rm bg}$
floated. The second and third method give comparable shifts in ${\hat p}^*(\Bz)$
and we take them as systematic uncertainty for the background-function; the
shift in the first method is 3 times smaller. Variations of the fit binning from
the nominal 15\mevc width to 5, 10, and 20\mevc have a negligible influence. The
transformation from laboratory-frame momenta to cms momenta has negligible
influence, even by varying the applied boost by the five-fold of its \rms in
\pep2. The detector influence on the $\Delta m_\B$ bias is estimated by using
the MC results for the means $\hat\delta$ of $p^*-p^*_{\rm true}$ as estimators
for the uncertainties. The results are
${\hat\delta}(\Bz)-{\hat\delta}(\Bp)=(-0.14\pm 0.13)\mevc$ for the $\mu\mu$ and
$(0.25\pm 0.19)\mevc$ for the $ee$ mode. We conservatively use the sums of
central value and \rms of these results for the last-line entry in
\reftab{syst}.

\begin{table}
\caption{\label{tab:syst}%
Summary of the systematic uncertainties for the measurement of $\Delta p^*$ in
\mevc.}
\begin{center}
\begin{tabular}{lc@{\quad}c}
\toprule
& \tabT\tabB $\mu\mu$ & $ee$ \\
\hline
\tabT Signal Fit-Function   & 0.12 & 0.17 \\
\Bp Background Fit-Function\quad\quad\quad & 0.01 & 0.03 \\
\Bz Background Fit-Function & 0.25 & 0.16 \\
Histogram Binning           & 0.08 & 0.08 \\
\tabB Detector Bias         & 0.27 & 0.44 \\
\hline
\tabT\tabB Quadratic Sum    & 0.40 & 0.51 \\
\hline\hline
\end{tabular}
\end{center}
\end{table}

Adding all systematic uncertainties in quadrature and taking the larger of the
two estimates ($ee$) leads to
\begin{equation*}
\Delta p^* = (-5.5\pm 0.8\pm 0.5)\mevc\,.
\end{equation*}
Inserted into \refeq{eq3} and using 319\mevc and 5279\mevcc for the mean values
of \B momentum and mass, we obtain
\begin{equation}
\Delta m_\B = (+0.33\pm 0.05\pm 0.03)\mevcc\,.
\label{eq:eqf}
\end{equation}
Contributions to the systematic uncertainty (in \mevcc) come from $\Delta p^*$
($\pm 0.031$), the track-momentum uncertainty ($\pm 0.011$), and the
mean-$p^*$-method bias ($\pm 0.007$). The contributions from the uncertainties
on the \Y4S boost and the \B-meson mass are negligible.

The $\Delta m_\B$ result in \refeq{eqf} is compatible with the present world
average~\cite{PDGweb} of $(0.37\pm 0.24)\mevcc$ but the error is a factor of 4
smaller. The significance of $\Delta m_\B$ being positive exceeds the $5\sigma$
level. Inserting our $\Delta m_\B$ result into \refeq{eq1}, we obtain
$R_0=1.051\pm 0.009$. The measured value of $R$ is
$1.037\pm 0.028$~\cite{PDGweb}. Given the agreement between these two results,
we do not observe significant Coulomb or quark-structure
contributions~\cite{Dubynskiy:2007xw} to $R$.

\begin{acknowledgments}
We are grateful for the excellent luminosity and machine conditions
provided by our \pep2\ colleagues, and for the substantial dedicated effort from
the computing organizations that support \babar.
The collaborating institutions wish to thank 
SLAC for its support and kind hospitality. 
This work is supported by
DOE
and NSF (USA),
NSERC (Canada),
CEA and
CNRS-IN2P3
(France),
BMBF and DFG
(Germany),
INFN (Italy),
FOM (The Netherlands),
NFR (Norway),
MES (Russia),
MEC (Spain), and
STFC (United Kingdom). 
Individuals have received support from the
Marie Curie EIF (European Union) and
the A.~P.~Sloan Foundation.
\end{acknowledgments}


\begin{thebibliography}{11}
\bibitem[{use()}]{useCC}
\bibinfo{note}{The use of charge-conjugate states in this analysis is
  always implied.}
\end{thebibliography}
\end{document}